\documentclass[twocolumn,prl,showpacs,amsfonts,superscriptaddress,floatfix]{revtex4-1}
\usepackage{graphicx}
\usepackage{amsmath}

\newcommand{\nextnearsites}[1]{\langle\!\langle #1 \rangle\!\rangle}
\newcommand{\nearsites}[1]{\langle #1 \rangle}
\newcommand{\phantomdagger}{{\vphantom{\dagger}}}
\newcommand{\phantoml}{{\vphantom{l}}}

\newcommand{\eref}[1]{(\ref{#1})}
\newcommand{\fref}[1]{fig.~\ref{#1}}

\begin{document}
\title {
Giant magnetoresistance of edge current between fermion and spin topological systems}
\author{Igor O. Slieptsov and Igor N. Karnaukhov}
\affiliation{G.V. Kurdyumov Institute for Metal Physics, N.A.S. of Ukraine, 36 Vernadsky Boulevard, 03142 Kiev, Ukraine}

\begin{abstract} {A spin-$\frac{1}{2}$ subsystem conjoined along a cut with a subsystem of spinless fermions in the state of topological insulator is studied on a honeycomb lattice.  The model describes a junction between a 2D topological insulator and a 2D spin lattice with direction-dependent
exchange interactions in topologically trivial and nontrivial phase states. The model Hamiltonian of the complex system is solved exactly by reduction to free Majorana fermions in a static $\mathbb{Z}_2$ gauge field. In contrast to junctions between topologically trivial phases, the junction is defined by chiral edge states and direct interaction between them for topologically nontrivial phases. As a result of the boundary interaction between chiral edge modes, the edge junction is defined by the Chern numbers of the subsystems: such the gapless edge modes with the same (different) chirality switch on (off) the edge current between topological subsystems. The sign of the Chern number of spin subsystem is changed in an external magnetic field, thus the electric current strongly depends both on a direction and a value of an applied weak magnetic field. We have provided a detailed analysis of the edge current and demonstrate how to switch on (off) the electric current in the magnetic field.
}
\end{abstract} \maketitle

\section{Introduction}

The Kitaev honeycomb model~\cite{Kitaev1} has given a great boost to the investigation of topological states in magnetic systems. A gapped phase with a chiral Majorana edge state and non-Abelian anyonic excitations in a bulk are realized in the model. The ideas stated in ref.~\cite{Kitaev1} have interesting physical applications, ranging from an exact method of solution of the models generalized to a variety of other lattices~\cite{k1,k2} to a study of topological chiral quantum spin liquids with gapless chiral edges states 
and spin liquids with Fermi surface. In the model of ref.~\cite{Kitaev1} the gauge fields are constants of motion, this fact being the key to the exact solvability of the model. Physical realizations of the spin-$\frac{1}{2}$ Kitaev model have been proposed in optical lattice systems~\cite{1a}.

An interest in the topological systems is inspired by special properties of their boundaries~\cite{Kitaev2,b}.
Excitations in the bulk are gapped in topologically trivial and nontrivial phases (in a weak magnetic field) and at the same time boundary excitations are gapped (in topologically trivial phase) or chiral gapless (in topologically nontrivial phase). In contrast to the junction between subsystems in topologically trivial phases, the junction between topological phases is leaded to a strong interaction between chiral edge states on the boundary. As a result, edge modes are deformed due to the boundary interaction.
The unique properties of the surface are responsible for their exotic electromagnetic properties, and might be used to realize new topological phases hosting Majorana modes when brought into contact between different topological systems.

The aim of the paper is to demonstrate `the work' of the junction between subsystems with different nature (spin and fermion subsystems) in topologically nontrivial states. We present certain exact results for chiral edges modes that define the chiral current along the cut for an arbitrary value of a coupling constant. The junction considered is unique in the sense that it the first exact description of the junction between different subsystems in topologically nontrivial states. We propose a principal new realization of Kitaev's ideas considering a spin sublattice connected by a junction with a topological insulator (TI), in other words, a junction between topological spin and fermion insulators. As a model system, the topological Kondo lattice thus identifies a new class of spin-charge insulator.
 The experimental situations when the spin-charge coherent transport is observable in the quantum spin Hall regime of a two-dimensional TI~\cite{kane} is discussed in ref.~\cite{ex}.

\section{Model}

\begin{figure}[t]
\centering{\leavevmode}
\includegraphics[width=8cm]{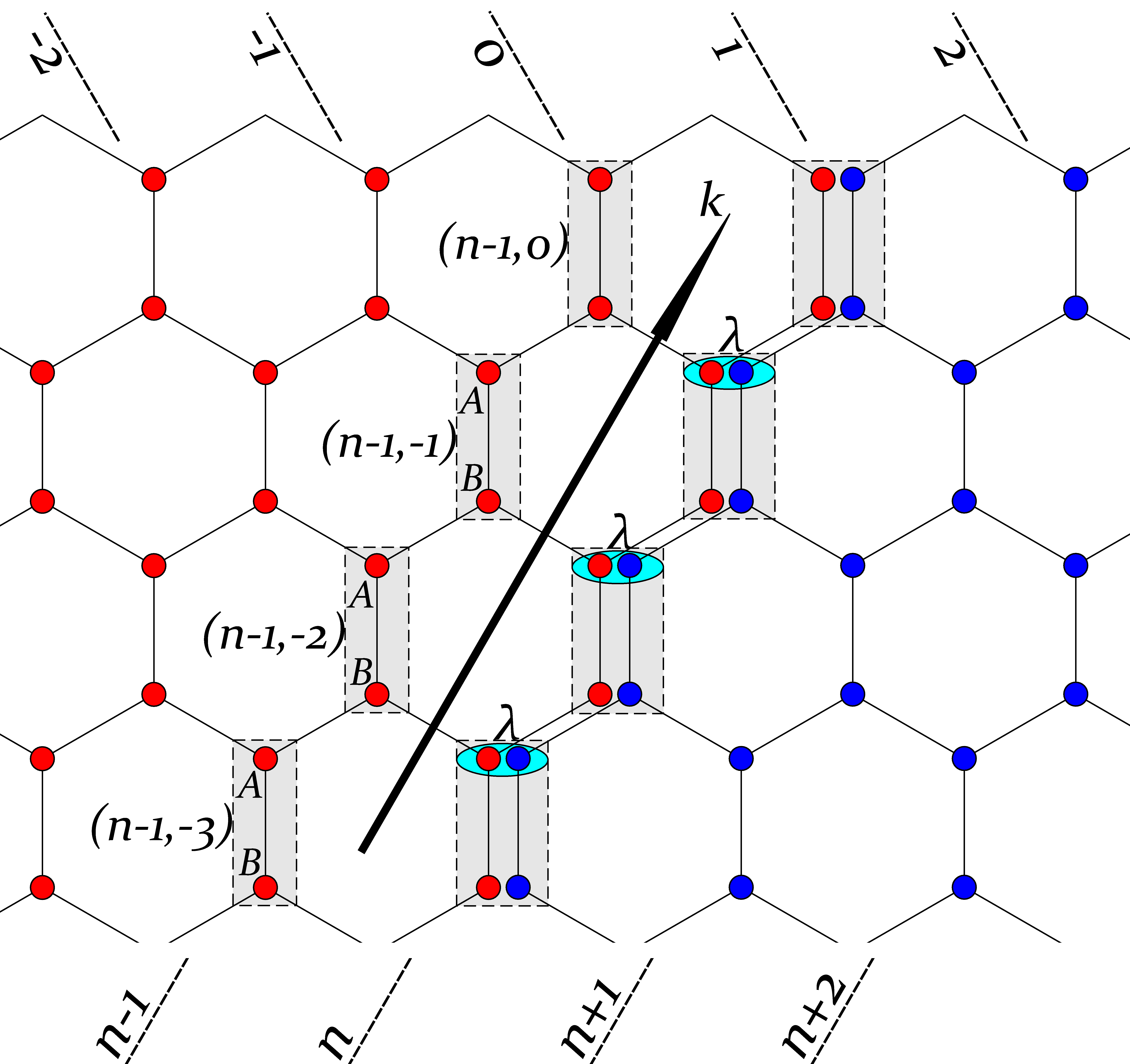}
\caption{(Color online)
Spin and fermion sublattices defined on a honeycomb lattice consisting of the spin (blue circles) and fermion (red circles) sites,
the unit cells highlighted by gray box,
$60^\circ$ coordinate system,
illustration of the zig-zag cut and the interaction term $\lambda$.
} \label{FigModel}
\end{figure}

We focus on a 2D model of spinless fermions in a topological state conjoined via a contact interaction with a spin-$\frac{1}{2}$ subsystem with direction-dependent exchange interactions. The model is defined on a honeycomb lattice with a spin-$\frac{1}{2}$ located at each site (see \fref{FigModel}). The total Hamiltonian ${\cal H}={\cal H}_{f}+{\cal H}_{s}+{\cal H}_{cut}$ describes fermion ${\cal H}_f$ and spin ${\cal H}_s$ subsystems conjoined via the boundary term ${\cal H}_{cut}$ along the cut.
We use the Kitaev model~\cite{Kitaev1} for the spin subsystem
\begin{equation*}
{\cal H}_{s} =
  I_x\sum_{x\mathrm{-links}} \sigma_{i}^x \sigma_{j}^x +
  I_y\sum_{y\mathrm{-links}} \sigma_{i}^y \sigma_{j}^y +
  I_z\sum_{z\mathrm{-links}} \sigma_{i}^z \sigma_{j}^z,
\end{equation*}
where $i$ is the nearest neighbor of $j$ connected by $x$-, $y$- or $z$-links, $\sigma_{j}^\gamma$ are the three Pauli operators at site $j$ and $I_{\gamma}$ is an exchange integral along a link $\gamma=x,y,z$.

For the sublattice of the spinless fermions we will use the following tight-binding model that takes into account hoppings between the nearest- and next-nearest-neighbors with equal hopping and pairing amplitudes
\begin{multline}
{\cal H}_{f} =
  - i t_1 \sum_{\nearsites{l,m}}    \left( a^\dagger_{l} a^{\phantomdagger}_{m\phantoml} + a_{l} a_{m\phantoml} \right) \\
  - i t_2 \sum_{\nextnearsites{l,m}}\left( a^\dagger_{l} a^{\phantomdagger}_{m\phantoml} + a_{l} a_{m\phantoml} \right) + h.c.,
\label{eq:Hf}
\end{multline}
where $a_{l}^\dagger$ and $a_{l}$ are the spinless creation and annihilation operators defined on the honeycomb fermion sublattice and satisfying the usual anticommutation relations, $t_1$ is an overlap integral of the nearest-neighbor hopping between neighbor sites $\nearsites{l,m}$ from~$l$ of the type~$A$ to~$m$ of the type~$B$ ($A$ and $B$ are the atoms of unit cell of the honeycomb lattice), $t_2$ is a clockwise (relatively honeycomb) the next-nearest-hopping between second-neighbors $\nextnearsites{l,m}$.

We have used the notion of Majorana fermions $d_l = a^\phantomdagger_{l} + a^\dagger_{l}$ and $g_l = \frac1i(a^\phantomdagger_{l}-a^\dagger_{l})$ for the Haldane Hamiltonian without a local staggered potential with $t_2\rightarrow t_2\exp(i \phi)$ in the special points $\phi = \pm \pi/2$~\cite{Hal} obtaining
\begin{equation*}
   {\cal H}_{f} = {\cal H}_f(d) = -it_1 \sum_{\nearsites{l,m}} d_l d_m -it_2 \sum_{\nextnearsites{l,m}} \nu_{lm} d_l d_m,
\end{equation*}
where $\nu_{ij}=\pm1$ stands for clockwise (anticlockwise) next-near-neighbour hopping inside corresponding plaquette.
In spite of the availability of the pairing terms in~\eref{eq:Hf}, the Fermi energy of electrons is well defined and the spectrum of the Hamiltonian ${\cal H}_{f}$ which 
is a half of the one-particle Haldane Hamiltonian
\begin{equation*}
  {\cal H}_{0} = {\cal H}_f(d) + {\cal H}_f(g) =
      - i t_1 \sum_{\nearsites{i,j}}\tilde{a}^\dagger_{l} \tilde{a}^\phantomdagger_{m}
      - i t_2 \sum_{\nextnearsites{i,j}} \nu^\phantomdagger_{lm} \tilde{a}^\dagger_{l} \tilde{a}^\phantomdagger_{m\phantoml}
\end{equation*}
which consists of two noninteracting ${\cal H}_f$-like parts of $d$ and $g$ Majorana fermions, $\tilde{a}_l$ is a fermion annihilating operator.

The term ${\cal H}_{cut}$ is governed by the interaction between local moments and itinerant spinless fermions located at the cut $\alpha$ (see \fref{FigModel})
\begin{equation}
{\cal H}_{cut}=\lambda\sum_{\alpha}(2a^\dagger_{\alpha}a^\phantomdagger_{\alpha}-1)\sigma^y_{\alpha},
\label{eq:Hint}
\end{equation}
where $\lambda$ is the coupling parameter, $\alpha$ denotes the boundary sites connected by two links ($z$- and $x$- links) along the cut. We consider only zig-zag cut along the direction shown in \fref{FigModel} with black arrow whose na\"ive cartesian coordinates are $(1/2,\sqrt3/2)$. The Hamiltonian~\eref{eq:Hint} defines the junction between subsystems along the zig-zag cut with broken $y$-links in sites on the cut.

The energy of external magnetic field $\vec{H}$ can be added to the total Hamiltonian of the system ${\cal H}_h = -\vec{H}\sum_{j}\vec{\sigma}_j$, where an external magnetic field acts on all spins.
The system is stable due to a time reversal (TR) symmetry, the energy spectrum of spinless fermions and Majorana fermions can not open gaps without breaking TR symmetry. In the case of noninteracting subsystems, an external magnetic field in the Kitaev model~\cite{Kitaev1} and hopping between the next-nearest-neighbor fermions in the Haldane model~\cite{Hal} break TR symmetry of the corresponding subsystem. The boundary term~\eref{eq:Hint} breaks TR symmetry of the total Hamiltonian~${\cal H}$, but does not stabilize the spin subsystem; being the boundary term, it only `deforms' the edge states. Below we consider an effect of the boundary interaction on edge states' structure in the charge and spin sectors.

\section{Topological edge current}

We follow Kitaev~\cite{Kitaev1} using the Majorana fermion representation
of spin-$\frac{1}{2}$ operators within an enlarged Hilbert space.
The spins are located at the vertices connected by three types of links, each link is
associated with a type of the interaction between nearest-neighbor
spins~\cite{Kitaev1}. The model is exactly solvable in the absence of an external magnetic field,
the Hamiltonian can be diagonalized
using a presentation of the Pauli operators $\sigma^\gamma_j=ib^\gamma_jc_j$ in terms of Majorana
fermions $b^\gamma_j$ and $c_j$ satisfying a Clifford algebra:
$\{c_i,c_j\}=2\delta_{ij}$, $\{b^\gamma_i,b^{\gamma'}_j\}=2\delta_{\gamma\gamma'}\delta_{ij}$, $\{c_i,b^\gamma_j\}=0$.

The total Hamiltonian can thus be rewritten in the following form
\begin{multline}
{\cal H} = -\frac{i}{2} \sum_{\gamma=x,y,z}\sum_{\nearsites{i,j}} A_{ij}^\gamma c_i c_j- it_1 \sum_{\nearsites{l,m}} d_l d_m
  \\-it_2 \sum_{\nextnearsites{l,m}} \nu_{lm} d_l d_m
  + \lambda\sum_{\alpha} g_{\alpha} d_{\alpha} c_\alpha b_\alpha^y,
\end{multline}
where the matrix $A$ consists of $A_{ij}^\gamma = 2I_\gamma u_{ij}^{\gamma\gamma}$ for the directed links $\gamma=x,y,z$.

In the absence of an external magnetic field, the operators $u^{\gamma \gamma}_{ij}=ib^\gamma_i b^\gamma_j$ and $u^{y}_{\alpha j}=ib^y_\alpha d_j$ commute with the Hamiltonian, they are constants of motion with eigenvalues $\pm1$, the model has an exact solution. The state with vortex-free field configuration is favored. Numerical calculations show that the interaction (along the zig-zag cut) between subsystems does not destroy the uniform flux (the ground state) sector. Let us simplify the problem and consider the magnetic subsystem with an isotropic exchange interaction $I_x=I_y=I_z=I=1$.

\begin{figure}[tbp]
\centering{\leavevmode}
\includegraphics[width=4cm]{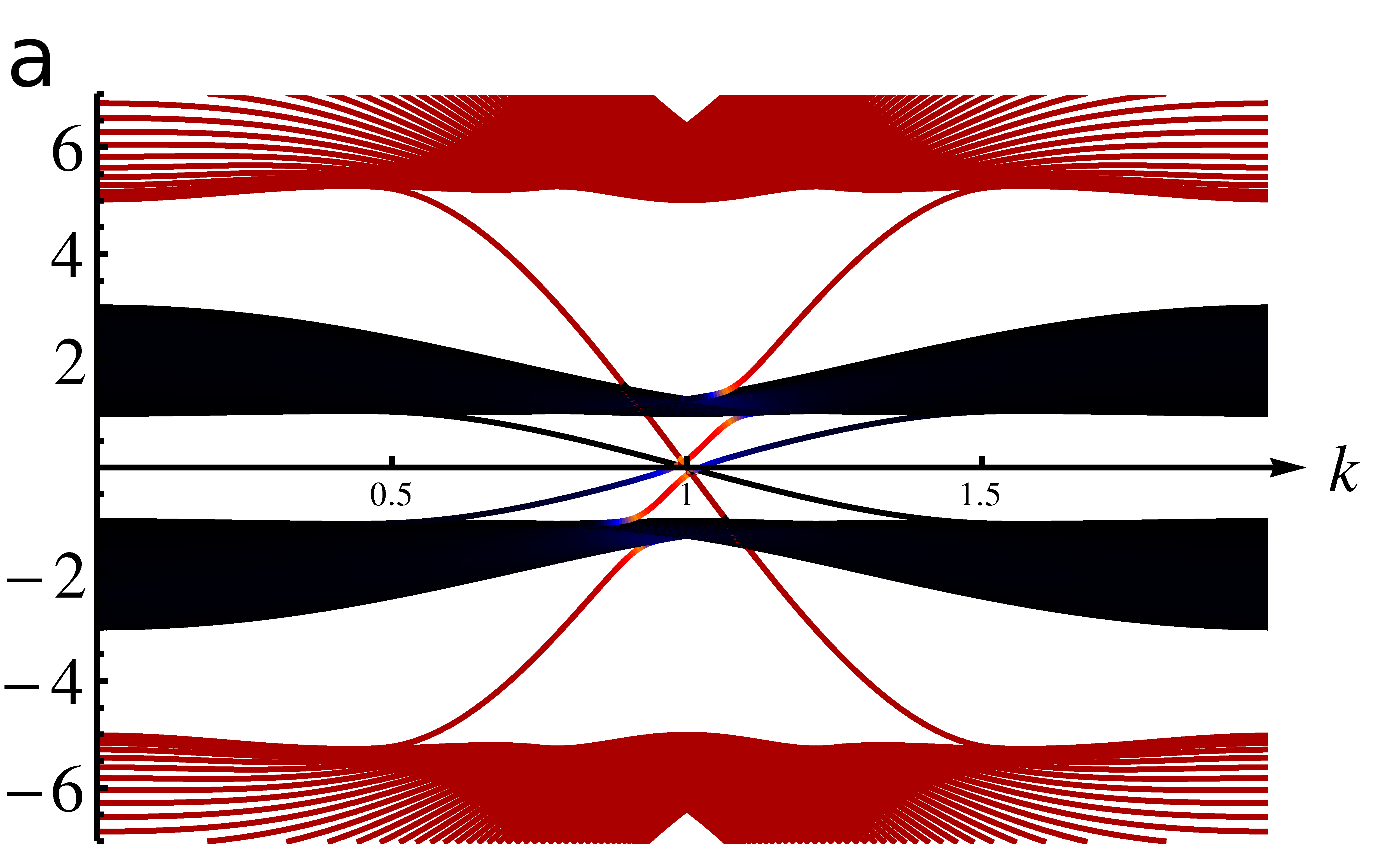}
\includegraphics[width=4cm]{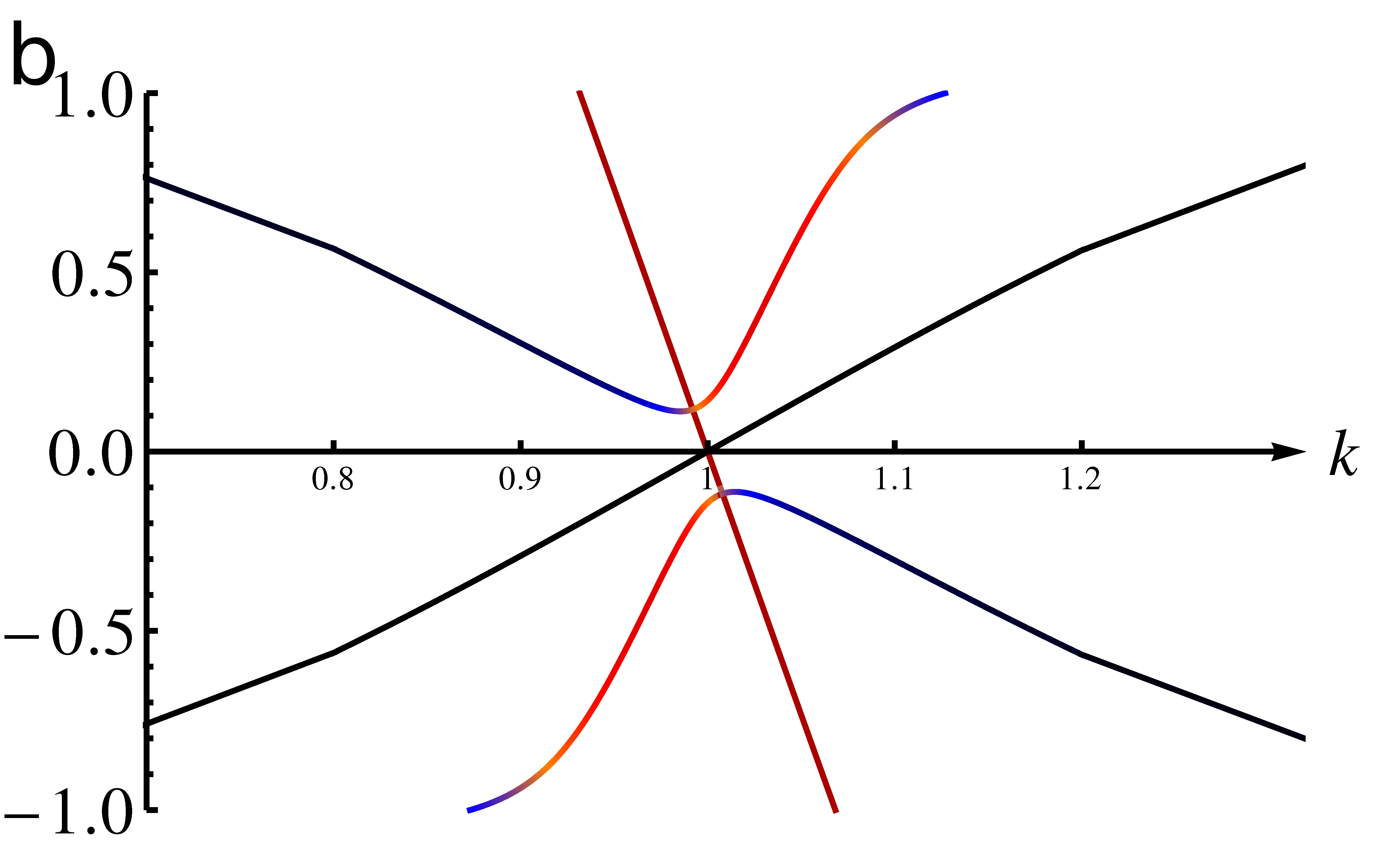}
\includegraphics[width=4cm]{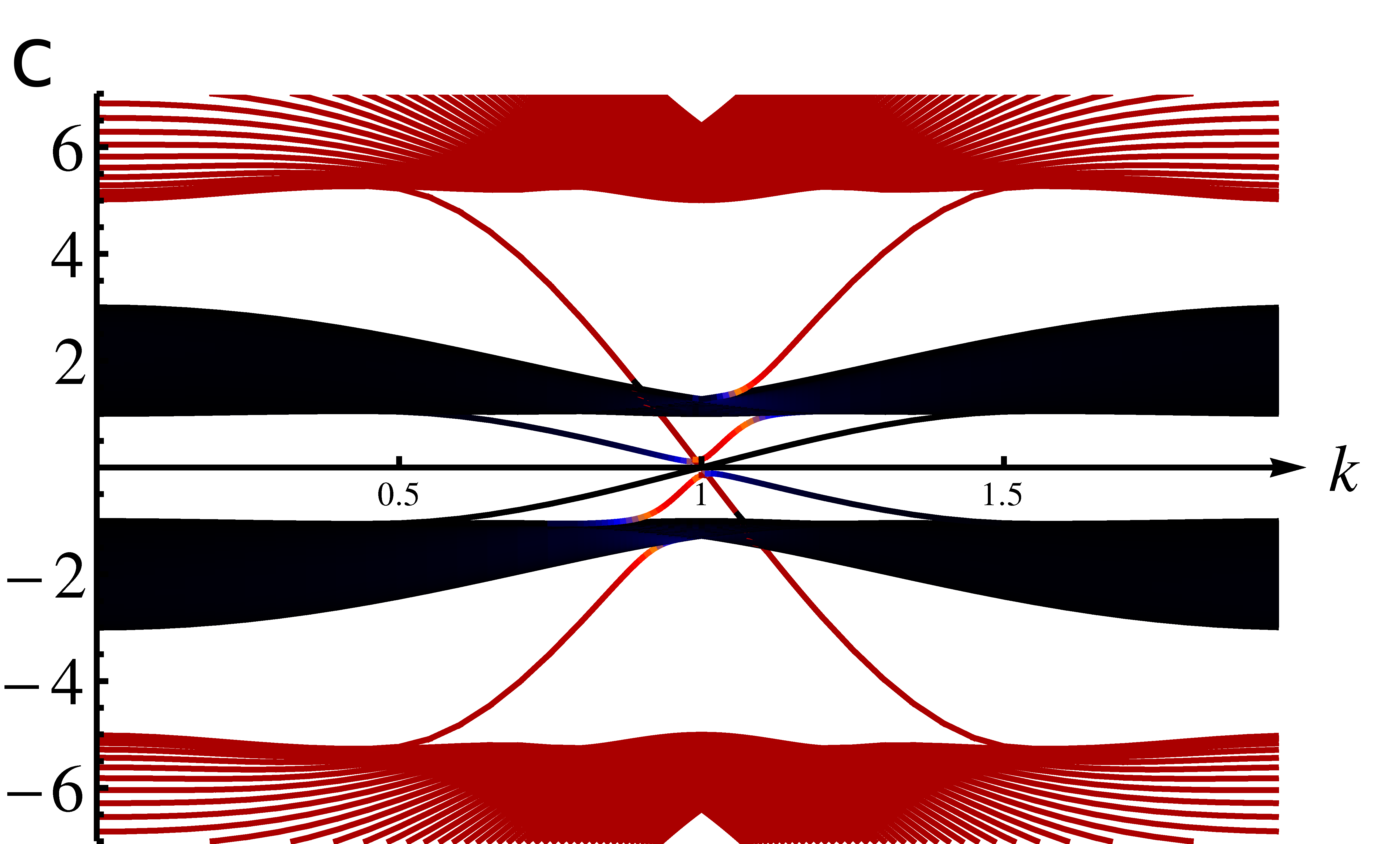}
\includegraphics[width=4cm]{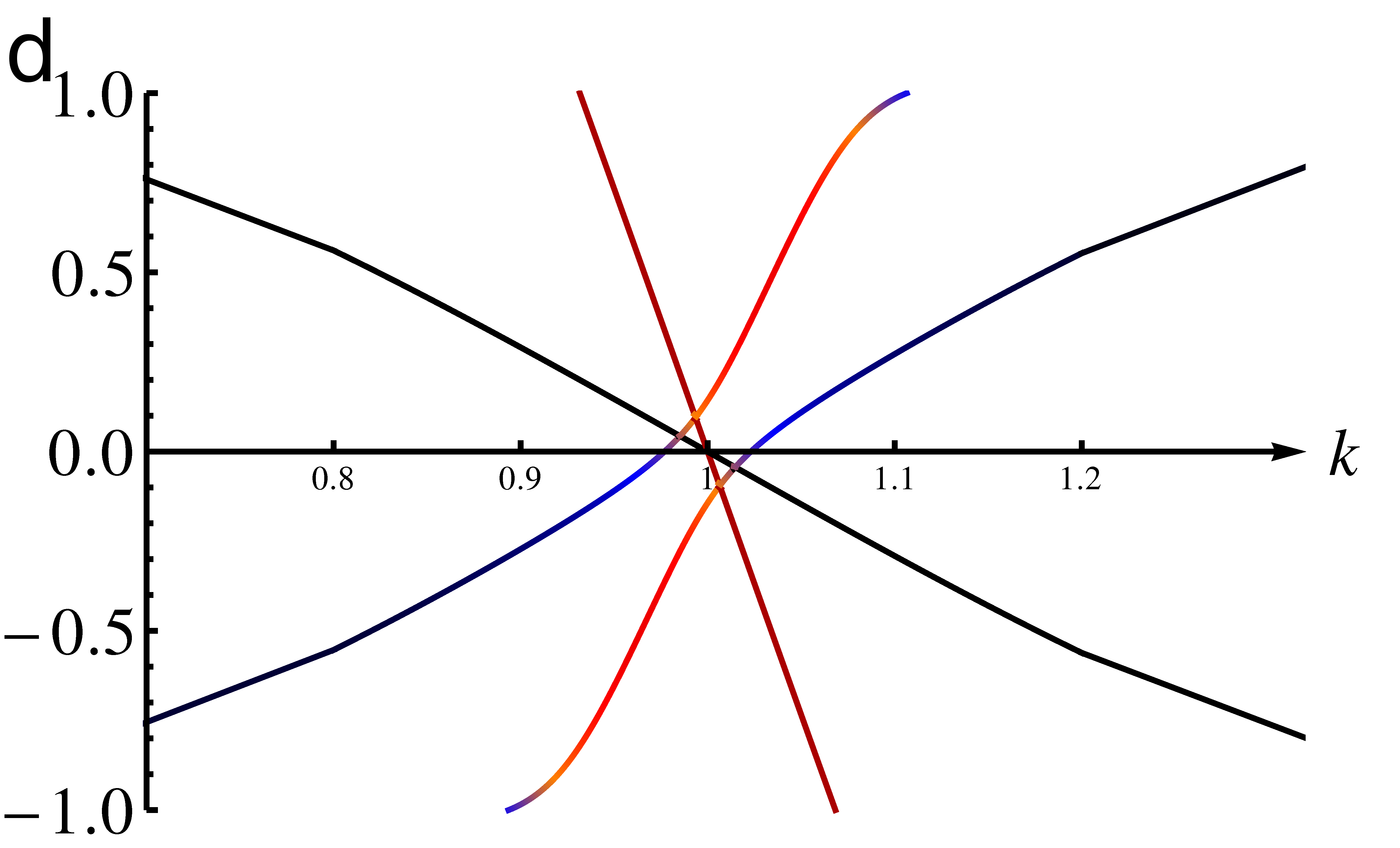}
\caption{(Color online)
  Dispersion $\varepsilon(k)$ of the system calculated on two cylinders with open boundary conditions, a junction for zig-zag boundaries with $t_1=5$, $t_2=1$, $I=1$, $\lambda=0.5$
  a) $\varepsilon(k)$ of bulk and edge modes along the zig-zag direction for $h=0.2$ b) the edge modes for $h = 0.2$,
  c) $\varepsilon(k)$ for $h = -0.2$, d) the low energy energy levels $h=-0.2$.
  Pure fermionic states are marked with red color, pure spin states are marked with black, mixed states are shown by gradient through orange (fermion part dominating) to blue (spin part dominating).
} \label{Edge}
\end{figure}

Consider the coupled subsystems in the presence of a magnetic field counted with perturbation theory according to Kitaev's approach, an effective magnetic field~$h$ is defined as $H_x H_y H_z/I^2\sim H^3/I^2$~\cite{Kitaev1}. We solve eigenvalue problem numerically and calculate energies of excitations in the sectors with uniform flux. The fermion subsystem is TI for arbitrary values of $t_2$~\cite{k1,Hal}, whereas the ground-state phase diagram of the spin system is defined by the values of exchange integrals. In the case of an isotropic exchange interaction the phase state of the Kitaev model~\cite{Kitaev1} is topologically nontrivial one with Chern number~1. Spin and fermion excitations with chiral edge modes for the zig-zag cut are shown in \fref{Edge} for different topological states of spin subsystem. The sign of the Chern number of~TI varies with the sign of~$t_2$~\cite{Hal}, whereas the sign of the Chern number of spin subsytem is defined by the sign of an external magnetic field~\cite{Kitaev1}.

\

The structure of edge modes depends on the Chern numbers of noninteracting subsystems: due to the boundary interaction the chiral gapless edge modes are `deformed' into gaped modes in the case of different Chern numbers (see \fref{Edge}(b)) and into gapless modes for the same Chern numbers (see \fref{Edge}(d)). In contrast to ref.~\cite{kane}, the chirality of the edge modes is conserved.
The chiral current along the cut is equal to zero (currentless state) in the first case, a dissipativeness current along the cut (current state) is realized in the second case. The Chern number of the spin subsystem depends on the direction of a magnetic field, therefore the current state of the junction is defined by an applied magnetic field. In order to change the value of the current along the cut between subsystems, we can change the direction of the magnetic field, in other words, an electric current along the cut depends anomalously on an applied magnetic field.

\begin{figure}[tbp]
\centering{\leavevmode}
\includegraphics[width=4.2cm]{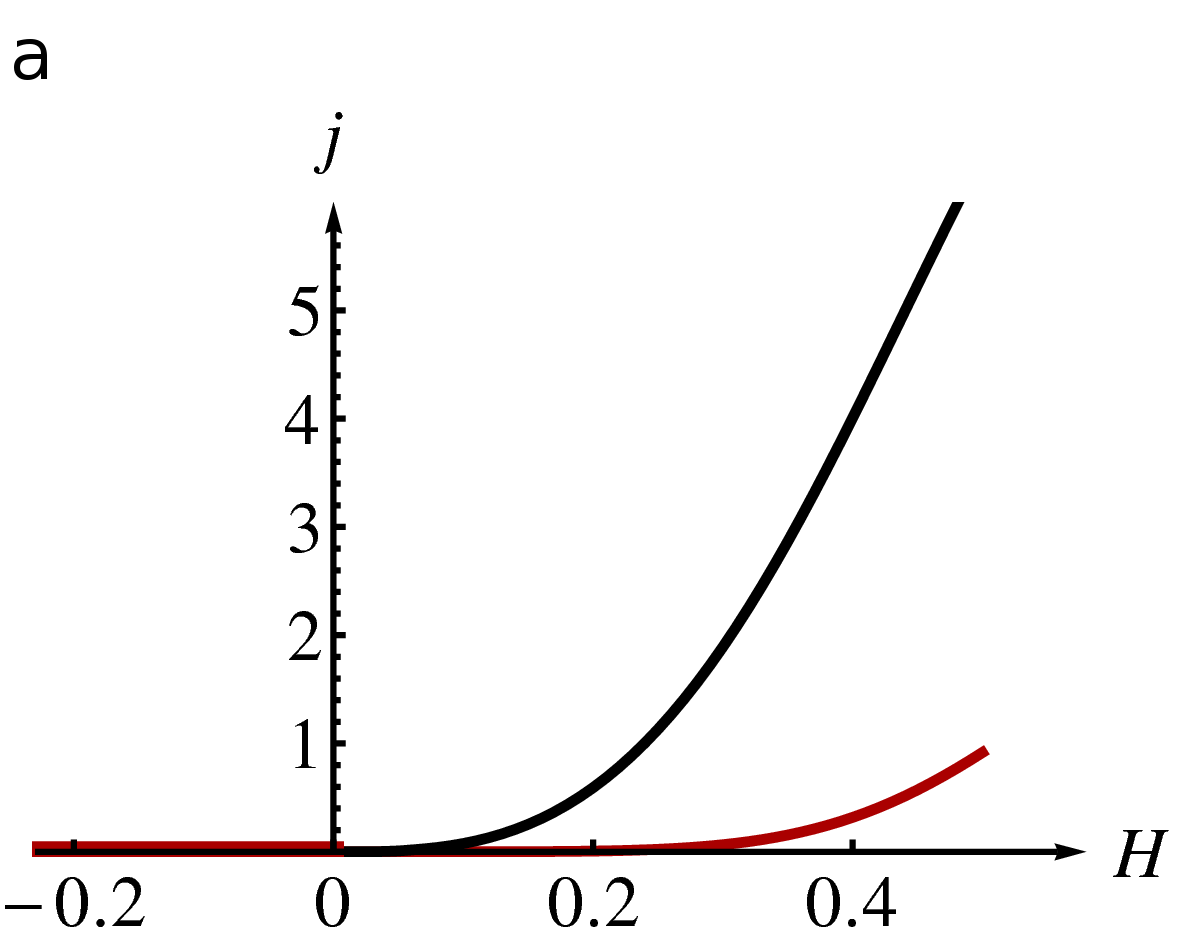}\includegraphics[width=4.2cm]{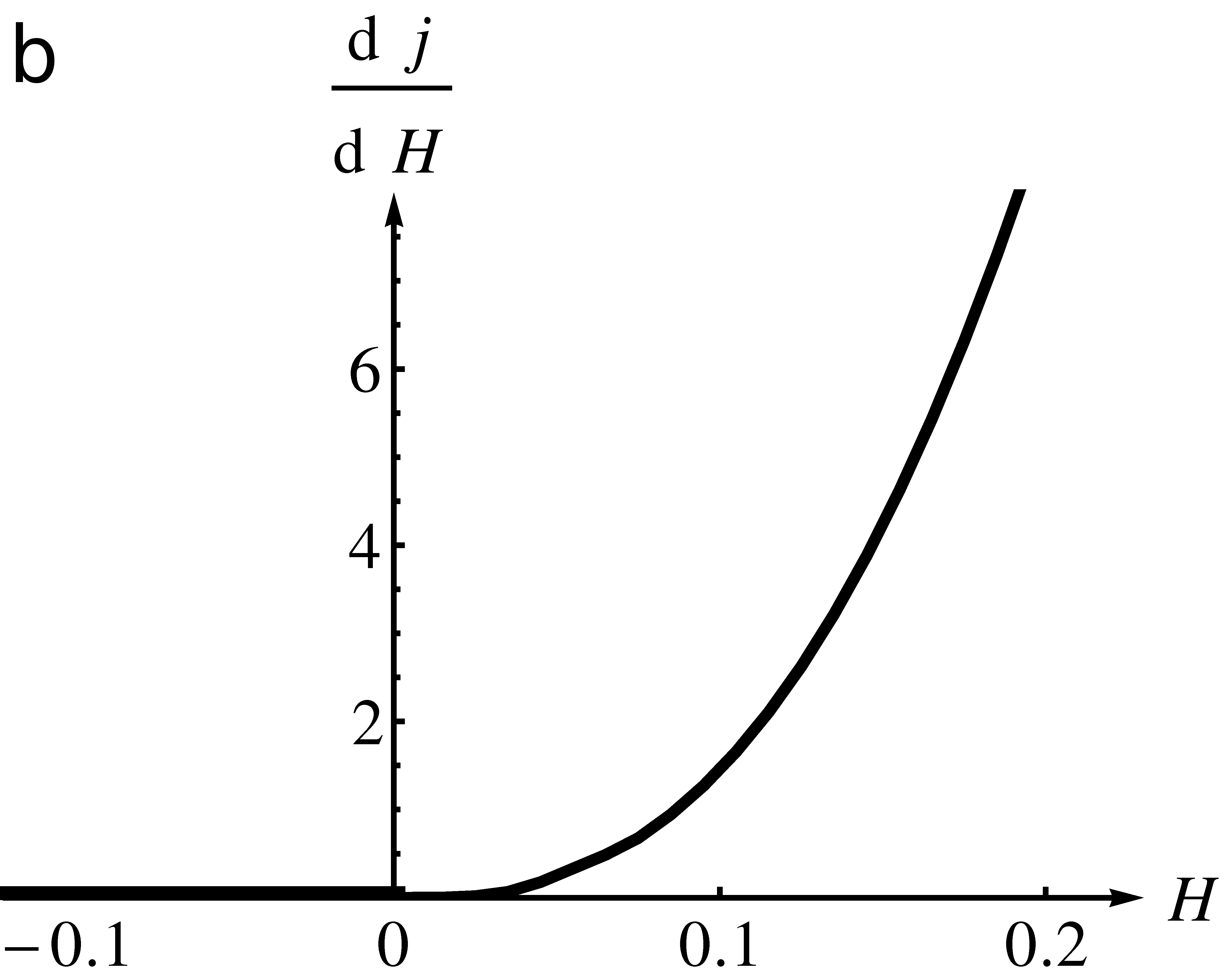}
\includegraphics[width=4.2cm]{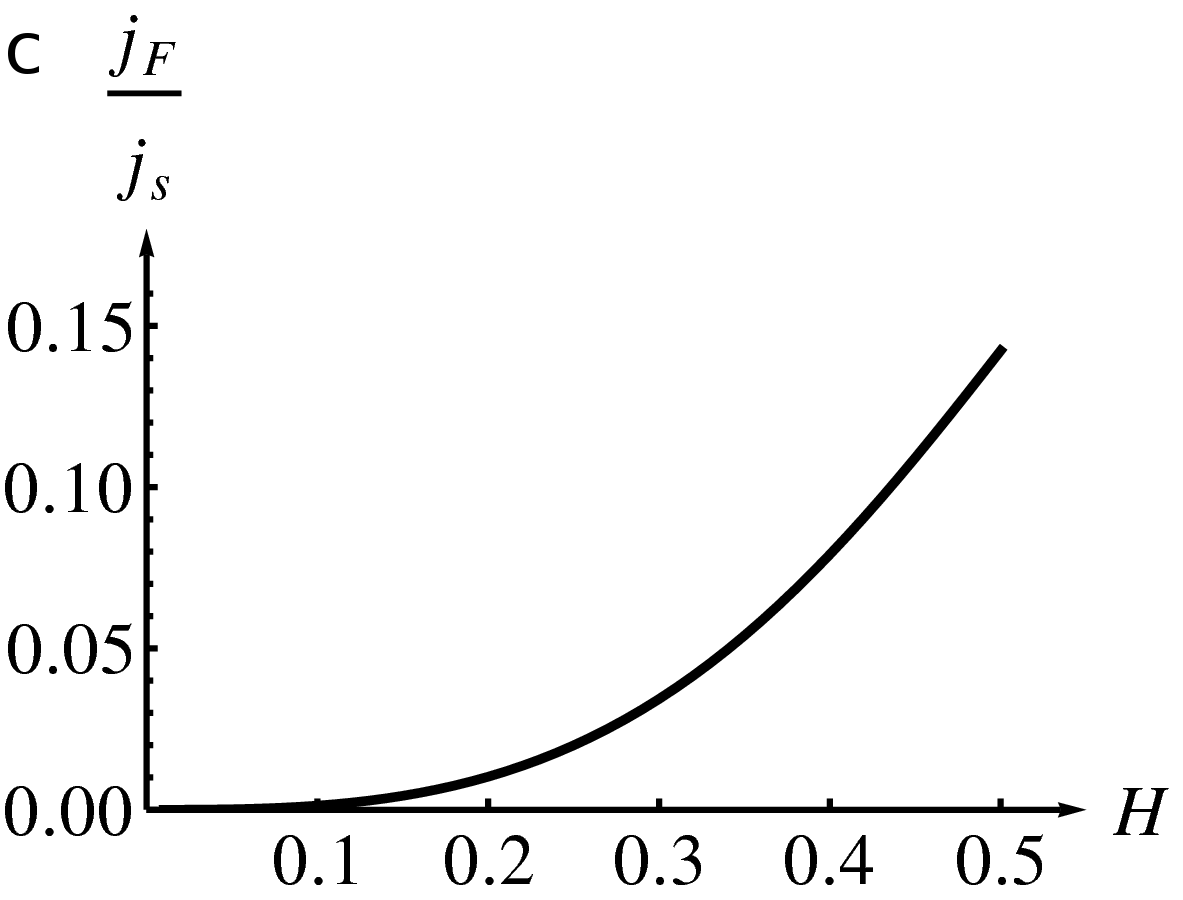}\includegraphics[width=4.3cm]{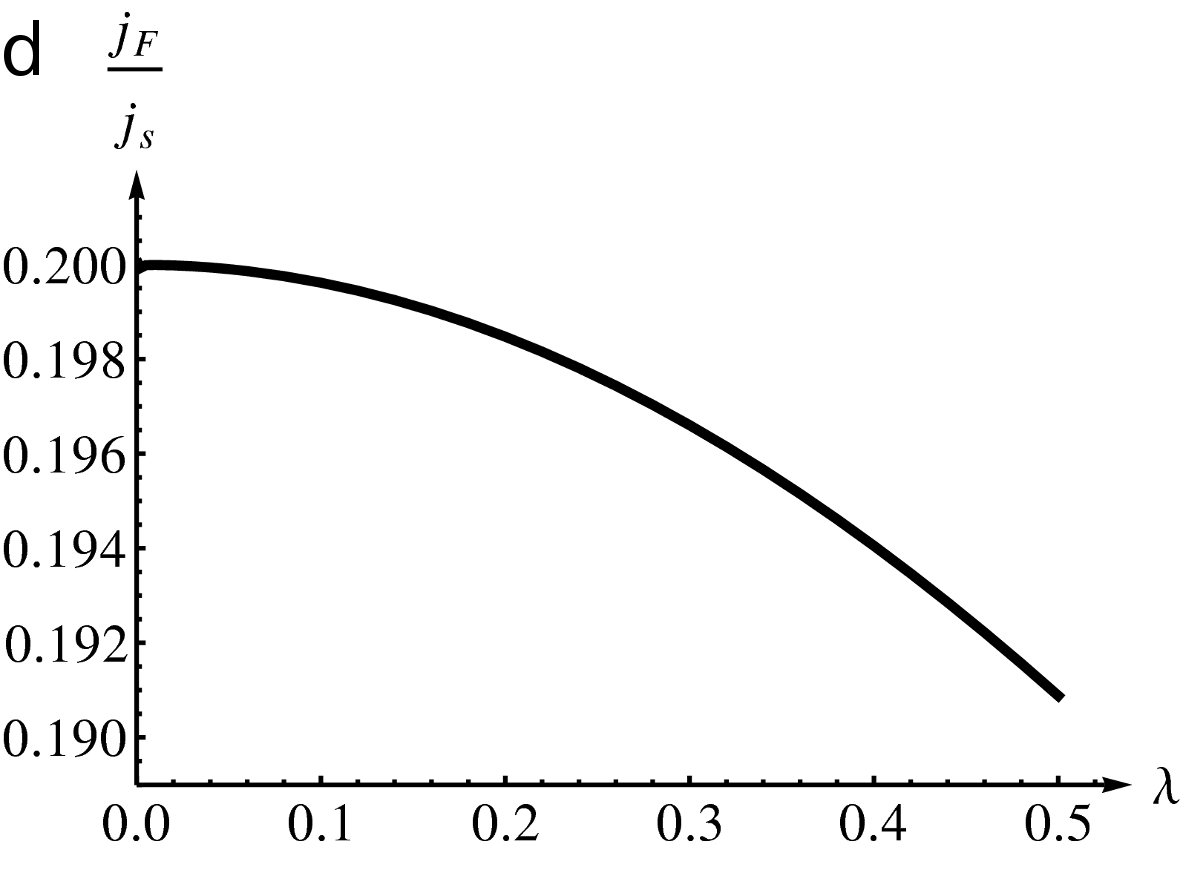}
\caption{(Color online)
  Fermionic (red curve) and spin (black one) currents along the cut and their ratio depending on model parameters $\lambda$ and $H$
  for $t_1=5$, $I=1$, $t_2=1$:
  a) currents for $\lambda=0.2$ with $H$ varying,
  b) derivative $R^{-1} = \frac{\partial j}{\partial H}$ depending on $H$,
  c) currents ratio depending on $H$ for $\lambda=0.2$,
  d) currents ratio depending on $\lambda$ for $h=0.2$.
} \label{Currents}
\end{figure}

Due to the hybridization of the edge modes, pure charge and spin chiral currents are not realized on the boundary, spin and charge currents are strongly correlated. We have calculated partial (spin and charge) currents as functions of the parameter of the interaction and the value of magnetic field, and displayed them in \fref{Currents}. The charge edge mode is high-energetic one, therefore the spin current dominates in low energy region (see \fref{Currents}). A main result is presented in \fref{Currents}(a) and \fref{Currents}(b) where the charge current $j_F$ and magnetoresistance $R=(\frac{\partial j_F}{\partial H})^{-1}$ calculated as a function of magnetic field. \fref{Currents}a and \fref{Currents}(b) illustrate the results obtained before in \fref{Edge} and show an anomalous dependence of chiral edge charge current on value and a sing of magnetic field. Low-temperature magnetoresistance of one-dimensional surface transport has been observed in topological Kondo insulator $SmB_6$~\cite{c}.

\section{Conclusions}

We have considered a 2D complex topological system~--- spin subsystem (in the framework of the Kitaev model) and TI (in the framework of the Haldane model in special points without a staggered potential) which are in contact along the boundary and calculated the chiral current along the boundary. We emphasize the fact that results obtained are exact, the models proposed have exact solution, therefore all the kinetic processes are dissipativeness. We have shown that the interaction on the cut leads to a strong hybridization of spin and charge edge modes and changes cardinally the transport of spin and charge in gapless edge states.
We have calculated the structure of chiral edge modes in topologically nontrivial phases of subsystems and partial edge currents. Due to the hybridization on the boundary, the gapless edge modes with different chirality are attracted and connected to one gapped edge mode. At the same time  the gapless edge modes with the same chirality are repulsed conserving gapless states. The state along the boundary is currentless in the first case and there is a current in the second case, the transition from one to another state is realized in weak magnetic field by changing a direction of an applied magnetic field. We have shown that topological properties of interacting systems with different nature (charge and spin) can define anomalous magnetic resistivity chiral edge current and lies to giant magnetoresistance effect.
It would be of interest to find materials which exhibit this effect, as well as possible 3D generalizations.

\end{document}